# The American Physical Society's Defense of Human Rights

By Edward Gerjuoy

*The author is a Professor of Physics Emeritus, Dept. of Physics, 100 Allen Hall, University of Pittsburgh, Pittsburgh, Pa. 15260, gerjuoy@pitt.edu*

This article describes the history of the American Physical Society (APS)'s involvement in the defense of human rights (HR) [1]. Until 1976 the APS, organized in 1899, had no formal mechanisms for engaging in HR activities. This is not to say that prior to then the community of American physicists was indifferent to governmental actions, by foreign governments or the US, which infringed on human rights. In 1938, for instance, many American physicists, including Nobel laureate Robert Millikan, signed a manifesto castigating Nazi racial theories [2], though no formal APS action was taken.

As Harry Lustig discusses in his history of the Society's first hundred years [3], the Society's failure to get formally involved in HR matters until 1976 can be ascribed to the lingering influences of the Society's founders, who believed the Society should remain aloof from public affairs not directly related to physics as a profession, a belief which was not unreasonable and no secret. Thus here [4] is what the New York Times had to say in 1948, when the APS took what for it at the time was the unusual action of openly criticizing the House Un-American Activities Committee for its attacks on Edward Condon, who was Director of the National Bureau of Standards and only two years previously had been the APS President:

> The American Physical Society, in a move unprecedented for an organization devoted exclusively to the affairs of pure science, entered the field of politics yesterday with a letter vigorously assailing the actions of the House Un-American Activities Committee in reference to Dr. Edward U. Condon... . The distinction between this message and those from other organizations lies in the fact that the American Physical Society prides itself on its aloofness from all matters except the intricacies of pure physics.

As World War II faded into the past, however, and as Vietnam War opposition grew in academic circles, more and more APS members, especially younger ones, pressed the Society to become seriously involved in public affairs. Such pressures first bore fruit in 1972 with the organization of the APS Forum on Physics and Society, the Society's first Forum. Three years later, in 1975, the year the Vietnam War ended, the APS created the first of the Society's so-called "outreach" committees, the Panel on Public Affairs (POPA) charged with advising the APS Council on public affairs issues.

## Creation of CIFS and its Pre-1980 HR Efforts

POPA soon conclude that public affairs issues of APS concern had to include HR violations the world over. POPA therefore set up a subcommittee to advise POPA, and ultimately the APS, on matters falling under the rubric of international human rights violations. By early 1976 this subcommittee had begun to function, named the Committee on International Freedom of Scientists (CIFS). A November 1976 *Physics Today* article [5] about CIFS' early activities described its formation:

> The subcommittee of POPA was established in response to the questions by The American Physical Society membership about scientists in Eastern Europe and South America whose rights and freedoms have been curtailed. The subcommittee was charged to provide facts and information for consideration by POPA and the Council, and to suggest constructive actions that might be taken by the APS Council.

Note that the original CIFS charge is consistent with the present CIFS charge:

> This Committee is responsible for monitoring concerns regarding human rights for scientists throughout the world. It apprises the President, the Board and Council of problems encountered by scientists in the pursuit of their scientific interests or in effecting satisfactory communication with other scientists and may recommend to the President and Council appropriate courses of action designed to alleviate such problems.

CIFS actions are therefore not restricted to matters affecting the APS and its physicist members, nor even to matters affecting physicists worldwide whether APS members or not. Instead CIFS is affirmatively charged to monitor concerns regarding the human rights of scientists, not merely physicists, throughout the world.

Yet POPA and CIFS were not established without opposition. As John Parmentola, a former POPA Executive Director, has written [7]:

> There was considerable controversy at the time between those advocating a very active role for the APS about speaking out and taking actions to ease the repression of scientists in the Soviet Union…and in South America, and those who felt this was "too political" for the APS. Professor C. S. Wu (the 1975 President of the APS) was quite enthusiastic about the establishment of CIFS, but others on the APS Council and even in the broader physics community were rather apprehensive about such a role for the APS. One has to appreciate that at the time, the APS was a very conservative body and the idea of involving the APS in international human rights activities was a very radical departure from its traditional role of publishing journals, organizing meetings, etc.

Despite this opposition, the proponents of establishing CIFS, and of using CIFS to vigorously support HR, carried the day. Moreover the APS has been willing to expend considerable funds and staff time in support of CIFS activities. Two APS staff persons, for instance, spend approximately 25% of their time working with CIFS and on CIFS-related issues. The APS provides travel support for these two staff members, and for nine CIFS members, to attend two Committee meetings a year. Furthermore, the APS administers the $10,000 Sakharov Prize, given every other year, "to recognize outstanding leadership and/or achievements of scientists in upholding human rights."

Soon after CIFS was formed its HR activities became both extensive and impressive. By May 1978, the APS Council had become so convinced of the importance of its human rights activities that it published a "Statement of Principles for the American Physical Society Activities With Regard to Human Rights" [9], whose opening paragraph stated that the APS activities in the area of human rights of scientists reflect the APS's conviction that science and scientific activity are important for the dignity of man and the future of civilization, and that interference with science anywhere is potentially harmful to all mankind and to society everywhere.

A report [10] by John Parmentola summarized CIFS activities from the date of its formation through September 1979. The main activity in those years was continued letter writing in support of scientists, almost all of whom were Russian. Despite the fact that Stalin was long since gone, in 1979 the Russian government still was continuing to deprive citizens, including scientists, of basic human rights. The Russian scientists whom CIFS supported between 1976 and 1979 mostly were so-called refuseniks [11], principally Jews who had been refused the right to emigrate and simultaneously had lost their employment. Parmentola's 1979 report lists the names of 21 refuseniks whom CIFS had been supporting. But writing letters was not the only CIFS activity in support of refuseniks. CIFS sought to seize whatever other opportunities to alleviate refusenik treatment came up. Thus as Parmentola relates, in April 1979 APS President Lewis Branscomb, at CIFS urging, wrote a Congressional delegation recommending that members of the delegation meet with members of the refusenik community during the delegation's planned visit to the Soviet Union. Moreover the delegation replied to Branscomb, indicating that members of the delegation not only actually met with refuseniks, but also raised problems that refusniks faced directly with Soviet authorities.

During those years CIFS also found itself supporting scientists, both Soviet and non-Soviet, who had suffered far worse human rights deprivations than Soviet refuseniks. For example, in 1977 APS President George Pake wrote letters to President Marcos concerning the imprisoned Philippine physicist Roger Posadas, and to President Ceaucescu concerning two Rumanian physicists whose freedom to pursue their profession had been restricted.

This discussion of CIFS 1976-79 activities closes with a few words about four serious HR deprivation cases, two involving Argentine scientists and two involving Russian scientists. In 1978, when a military junta was exercising power, two physicists residing in Argentina simply disappeared. One, Alfredo Giorgi, was the head of a plastics research laboratory in Buenos Aires, who was taken from his office by army and police officers and never heard from again. The second, Daniel Bendersky, was a graduate student studying nuclear physics. In December 1978 APS President Norman Ramsey wrote to the President of Argentina inquiring about Bendersky, only to be told that Bendersky's whereabouts were unknown.

Russian scientists who received human rights deprivations more serious than the refuseniks included physicist Yuri Orlov [12] and applied mathematician Natan Sharansky [13]. In 1978, both were sentenced to long prison terms for HR activities the government deemed anti-Soviet. Orlov had acquired the enmity of the Soviet authorities by, among other things, founding the Moscow Helsinki Group to monitor Soviet adherence to the Helsinki human rights accords. Sharansky was accused of passing to Western democracies the names of over 1300 refuseniks. In fact, the human rights organization SOS took its name from Sakharov, Orlov and Sharansky.

Both Orlov and Sharansky were freed in 1986, shortly after Gorbachev came into power. Orlov emigrated to the United States, where he became a member of the Cornell physics department. He also continued his HR activities, to such good effect that in 2005 he was named the first recipient of the Andrei Sakharov Prize, awarded by the APS "to recognize outstanding leadership and/or achievements of scientists in upholding human rights." Sharansky emigrated to Israel after his release from prison. He became active in Israeli politics, co-founded an anti-Palestine political party, and became a member of the Israeli cabinet. He resigned from the cabinet in 2005 to protest plans to withdraw Israeli settlements from the Gaza Strip and the West Bank. The Wikipedia article about him [13] shows a photo of Sharansky and Vladimir Putin having a private meal together. As the grandmother of this article's writer would say, "Go figure".

## CIFS HR Activities in 1980

By the end of 1979 CIFS's activities were taking so large a fraction of POPA's time that in 1980, only four years after CIFS had begun its work, the APS Council split CIFS from its parent APS committee POPA and established CIFS as an independent committee with essentially the same charge it has today. As an independent committee CIFS continued the HR activities it had undertaken as a POPA subcommittee, but was able to draw on even larger APS resources. Thus the CIFS 1980 Annual Report to the APS Council [14] lists a larger variety of human rights activities than this article has previously described. These included:

- Preparation of a report on adherence to the Helsinki Final Act for the U.S. delegation to an international Madrid Conference, providing detailed information about Soviet treatment of refuseniks, as well as of scientists like Orlov who were imprisoned for political reasons.
- Issuance of a report for the use of anyone who wishes to become involved, as an individual, on behalf of scientists who have been deprived of their human rights.
- Getting the APS President to send letters, plus supplemental materials prepared by CIFS, to the presidents of physical societies in Europe, Canada, Japan and Australia, soliciting cooperation in human rights activities.
- Protesting the U.S. government's intervention into scientific conferences by, e.g., requiring foreign participants to sign pledges that they would not transmit to Soviet-bloc nationals information learned at the conference.
- Protesting to the Director of the Lawrence Livermore Laboratory about the Lab's disciplinary notice to lab physicist Hugh DeWitt [15], because DeWitt had submitted affidavits opposing the government's attempt to suppress publication by *Progressive Magazine* of an article on the H-bomb drawn from sources in the open literature.

Note that some of these activities involve protests against the U.S., rather than against foreign governments. Furthermore, 1980 was the first year in which the APS President intervened on behalf of Andrei Sakharov, who that year was sent to internal exile in the city of Gorky, a city off limits to foreigners.

## Post-1980 CIFS HR Activities

After 1980, the CIFS's activities have been similar to those described for 1980. On its website [16], the CIFS classifies its activities into five categories:

- Reviews cases involving reported violations of the human rights of scientists throughout the world;
- Advocates by writing letters on behalf of scientists to relevant agencies and/or governments;
- Investigates by seeking additional information from colleagues, agencies, and other sources;
- Supports scientists whose rights have been violated through email and other forms of contact;
- Educates colleagues in the APS and other agencies through articles in APS News, presentations, and informal discussion.
- 

Examples of such activities include:

- *Scheduling sessions on HR subjects at APS meetings. At the 1981 Annual Meeting in New York, for example, the Forum on Physics and Society sponsored a CIFS-organized session featuring talks: by an exiled Argentine newspaper editor, by a member of Moscow Helsinki Watch who had just emigrated to this country, and by Congressman George Brown of the House Subcommittee on Science, Research and Technology, whose marvelous speech quoted Sakharov.*
- *Offering free APS membership and/or journal subscriptions to victimized physicists. This is program began in 1979, but by 1983 had become so expensive that the APS decided to regularly approve half-member rates only. These half-cost subscriptions were made available to most third-world physicists and libraries. Moreover, oppressed scientists continued to receive free subscriptions via an APS-publicized program of seeking membership donations for such subscriptions. In 1985, for example, there were thirty free memberships of this special sort, including Yuri Orlov, then still serving his 1978 twelve year prison sentence for having organized a Moscow Chapter of Helsinki Watch, and Fang Lizhi for a year or so immediately after the 1989 Beijing Tiananmen Square massacre, when he evaded arrest by taking refuge in the United States Embassy until permitted to emigrate [18].*
- *Initiating and/or writing articles describing APS HR activities in Physics Today and other publications. Physics Today* has published numerous other articles on human rights subjects besides Congressman Brown's talk, including: a 1981 article, "Soviet Repression of Dissidents" [19], featuring a photograph of refusenik Victor Brailovsky [20]; a 1985 article [21] describing CIFS activities; and a September 1989 article [22] detailing CIFS activities on behalf of Tayseer Aruri, a West Bank Palestinian physicist imprisoned by Israel and threatened with deportation. Since 1995, when publication of the monthly APS News began, many APS News stories have discussed aspects of the Society's human rights efforts. In 2014 alone five issues carried stories detailing CIFS activities on behalf of Omid Kokabee, an Iranian citizen pursuing a physics Ph.D. at the University of Texas who, in 2011 on a winter break visit to his family, was sentenced to 10 years imprisonment for refusing to work on Iranian military projects [23]. The 2014 *APS News* stories also discussed attempts to secure the rehiring of the Russian scientist Alexander Gorsky, who in March had been fired from the Moscow Institute of Theoretical and Experimental Physics for giving an invited talk at a meeting at Stony Brook University without permission from the Russian government [25].
- *Sharing information and otherwise cooperating with non-APS groups seeking to defend human rights.* At a number of APS meetings, for instance, CIFS arranged for the Committee of Concerned Scientists (CCS) to set up a table where APS members could sign CCS-prepared petitions on behalf of various oppressed scientists such as Kokabee [24]. CIFS-furnished information about the exaggerations of the U.S. government's testimony against Los Alamos researcher Wen Ho Lee (see below) convinced Amnesty International to write the presiding judge in support of Lee. Another illustration of cooperative activity was this author's trip to the Soviet Union in 1981, under the joint sponsorship of the APS and various Councils for Soviet Jewry, with the express purposes of visiting around 40 refuseniks in Moscow and Leningrad, bearing gifts and publications from the trip's sponsors and reporting back to the APS and the Councils about the circumstances of refuseniks whose names had been furnished by CIFS and/or the Soviet Councils. Since the Soviet authorities' typical reason for refusing a refusenik permission to emigrate was possession of state secrets, it's curious that one refusenik visited on this trip, Lev Blitshtein, was not a scientist but had worked in a sausage factory.

These examples do not include illustrations of letter writing activities that continued those begun before CIFS was established as an independent Committee. Such letter writing included, for instance: in 1980, protesting the U.S. government's refusal to allow Soviet scientists to attend an unclassified conference organized by the American Vacuum Society; In 1983, protesting Israel's refusal to permit Palestinian physicists to teach in West Bank universities unless they signed a commitment against "terrorist activities"; in 1983-4, asking UNESCO to investigate and redress Soviet violations of Orlov's human rights; in 1987, protesting the Chilean government's firing of physicist Carlos Infante and other University of Chile faculty; in 1988, protesting the Chinese government's refusal to permit Fang Lizhi to travel to the U.S.; in 1993, inquiring about several professors dismissed from Ethiopia's Addis Ababa University for speaking out about a brutal suppression of a student demonstration; in 2000, decrying U.S. imprisonment of Wen Ho Lee without bail; and, since 2011, letters seeking the release from prison of Omid Kokabee, including in particular a letter to Vladimir Putin from APS President Malcolm Beasley [26].

## Small Committee Letter Writing

The preceding paragraph's illustrations of CIFS letter writing after becoming an independent APS Committee have not included many letters written under the auspices of its so-called "small committees." CIFS initiated this program during the years 1976-78, while still a subcommittee of POPA, copying a practice said to be developed by Amnesty International. Each small committee, consisting usually of three persons, "adopts" a single persecuted scientist and agrees to write said scientist and his/her family on a regular basis, whether or not there is evidence the letters are being received. Even if the letters are intercepted, they demonstrate that the victimized scientist is not forgotten by the outside world, thereby hopefully easing the scientist's treatment or at least deterring extreme persecutions like torture.

Of the many APS human rights activities, this has been one of the most successful yet least publicized—to APS members as well as to the general public. The program began with only a few committees, but the number of committees grew rapidly, so that it soon proved necessary for the program to have a "coordinator". For example, in 1983 there already were 63 small committees, writing to the same number of oppressed scientists; these 63 committees were composed of 97 individuals, implying that many of the small committee members had accepted the responsibility of writing to more than one victimized scientist.

By 1985 the number of small committees had increased to 84 with 167 members. Most small committee members were physicists, and just about all scientists. The APS, and indeed the entire world, owes a long overdue expression of gratitude to every one of those small committee members who essentially anonymously, without fanfare, regularly wrote so many letters of encouragement to so many HR victims, often with little expectation that the letters would reach their intended recipients. Heartfelt thanks also are owed the small committee coordinators, especially Julian Heicklen of Penn State University, Edward Stern of the University of Washington, and Bernard Feldman of the University of Missouri, each of whom was willing to undertake the important task of coordinating the small committees over an extended period, even though coordination required a considerable expenditure of time.

The number of small committees reached its maximum of 101 in 1986, but decreased fairly steadily thereafter. By 2000 the number had fallen to 10. In 2001, CIFS voted to terminate its small committee program, ending organized letter writing by APS members to human rights victims. Even if the small committee format has outlived its usefulness, it is regrettable that the APS has not retained some mechanism whereby regular communications to selected human rights victims and their families, serving the morale raising and related functions which have been described herein, can be efficiently initiated.

Of the 84 scientists supported by small committees in 1985, all but two were in the Soviet Union; the two non-Soviet scientists were Polish. This small committee singling out of Soviet scientists is easy to understand. By the 1970s the United States physics community had become well acquainted, personally as well as professionally, not only with the Soviet physics community but also with many other Soviet scientists; in those years the American physics community was far better acquainted with the Soviet scientific community than with any other scientific community living under a repressive regime, e.g., the Chinese scientific community. Thus the ruthless Soviet persecution of large numbers of scientists during the 1970s and 1980s, many merely for peacefully criticizing their government or for seeking to emigrate, drew the attention of many American physicists and even earned explicit recognition from the beginning, in the assertion that CIFS was established in response to questions by APS members "about scientists in Eastern Europe and South America whose rights and freedoms have been curtailed."

As the 1980s drew to a close, more and more previously persecuted Soviet scientists were released from prison and/or permitted to emigrate, with the result that the number of Soviet scientists requiring and/or actually receiving small committee support rapidly began to decrease. For instance in 1987 the number of small committees was only 77, down from 84 in 1985. Moreover, as the number of small committees serving persecuted Soviet scientists decreased, the number of small committees serving persecuted scientists of other nations began to increase, reflecting the growing awareness of human rights abuses worldwide. Thus in 1989, when the number of small committees had fallen to 62, two of those committees were supporting the Palestinian physicists Sami Kilani and Salman Salman, and a third was supporting the Cuban physicist Jorge Molina.

These just discussed small committee trends were accelerated by the 1991 collapse of the Soviet Union, as well as by the 1989 Tiananmen Square massacre, which greatly raised APS membership awareness of Chinese human rights violations. Accordingly, of the 12 new small committees started between November 1989 and March 1990, six were for Chinese physicists and another for a Palestinian physicist; only five were for refuseniks.

Indeed, of the ten aforementioned small committees still existing in the year 2000, shortly before the small committee program was dissolved, only one was devoted to a scientist victimized by the former Soviet Union or by one of its daughter republics. The other nine committees were supporting two Cuban scientists, two Chinese scientists, a Vietnamese, a Palestinian, an Israeli, a scientist from Myanmar, and an American. The ten-paragraph CIFS Annual Report for the year 2000 devotes only a single paragraph to the human rights violations of a Russian scientist, Alexander Nikitin. A single paragraph is devoted to the imprisonment of the Cuban physicist Felix Carcasses for trying to set up a human rights group in Cuba. A single paragraph also is devoted to the case of Kin-Yip Chun, a seismologist at the University of Toronto who apparently was denied tenure because of his Chinese ethnic origin.

# Wen Ho Lee

Most of the 2000 CIFS Annual Report is devoted to matters involving Los Alamos researcher Wen Ho Lee, whose human rights were violated by U.S. security personnel. The Lee case facts that this article is about to relate are from transcripts of judicial hearings and a fourteen page House of Representatives discussion of the case [27].

Lee is a native Taiwanese who received a mechanical engineering Ph.D. from Texas A&M and became a U.S. citizen. From 1980 until he was fired in 1999 he worked for the Los Alamos National Laboratory (LANL), primarily as a programmer. At the time LANL had two main computers, an "internal" computer which had no connection to the outside world and an "external" computer connected to the outside world; it could receive emails from anyone. All classified LANL material was supposed to be kept solely on the internal computer, so that there would be no way for anyone outside LANL to access LANL classified material. The government's case alleged that while he was employed Lee had downloaded highly classified files from the internal computer to the external computer. Lee did not deny downloading files from the internal computer to the external computer, but claimed they were not highly classified.

There is no doubt that Lee breached LANL security restrictions and deserved punishment. But LANL security personnel became fixated on the notion that Lee was spying for a foreign government (unnamed in the indictment but presumably China). They secretly followed his movements for nine months, apparently hoping to catch him contacting some foreign agents. After those nine months, during which they had not caught him doing anything reprehensible and had subjected him to lie detector tests (which he passed), Lee was arrested and indicted on 59 felony charges, many espionage charges punishable by life imprisonment.

Lee's indictment was followed by court hearings at which the government presented evidence that purportedly showed how important for nuclear bomb design were the documents Lee had downloaded. The government therefore insisted that Lee was too dangerous a spy to be released on bail. Lee's lawyers were unable to locate witnesses who could counteract the government's case. When Lee's lawyers asked the Court, if it denied bail, to at least let him avoid jail by allowing him to stay at home under house arrest with the house guarded so Lee could not escape, the government put on a security expert who testified Lee was so important a spy that it feared the unnamed foreign power would send in a helicopter carrying armed men who would kill the security people guarding the house and then would spirit Lee away to that foreign power's territory.

U.S. District Court judge James Parker ruled that Lee should be imprisoned. The government then incarcerated Lee, confining him to his cell for 23 hours a day. He was not allowed to have contacts with other prisoners, even during the one hour a day he was allowed out of his cell so he could exercise. When moved out of his cell he was shackled at the waist, wrists and ankles. He could not make telephone calls except to his attorneys. Aside from his attorneys he could be visited only by his immediate family, in visits limited to one hour a week, during which the conversations were in the presence of an FBI operative and conducted in English, not Chinese. He had no access to newspapers, TV or radio. His family was not permitted to send him books; he could only have books that were mailed to him directly by a bookseller. These restrictions eventually were eased.

When CIFS became aware of the Lee case it began efforts on his behalf. This writer found a highly respected former Los Alamos nuclear bomb designer, John Richter, who was willing to testify about the weaknesses in the government's case. Lee's attorneys were then able to obtain a new bail hearing, about nine months after his original hearing, before the same Judge Parker who originally had refused Lee bail. Richter's testimony demolished the government's case. Also by this date Lee's lawyers had been able to establish that before the first hearing the government had increased the secrecy classifications of the documents Lee had downloaded, that when Lee downloaded them they had the lowest secrecy level, barely above unclassified.

The government, aware that it really had no case, agreed to a compromise whereby all but one of the counts against Lee were dismissed; Lee agreed to plead guilty to that one count; the government in turn agreed that the penalty for the guilty plea should be time served, i.e., to the time Lee had already spent in prison but nothing more. Judge Parker accepted the compromise, thus finally making Lee a free man, but before closing the hearing he did something practically unheard of: he publicly apologized to Lee for the original denial of bail. Here is part of what he said:

> I believe you were terribly wronged by being held in custody pretrial in the Santa Fe County Detention Center under demeaning, unnecessarily punitive conditions…I am sad that I was induced in December to order your detention, since by the terms of the plea agreement that frees you today without conditions, it becomes clear that the Executive Branch now concedes, or should concede, that it was not necessary to confine you last December or at any time before your trial…I sincerely apologize to you, Dr. Lee, for the unfair manner you were held in custody by the Executive Branch. The Lee case [28] provides another, indeed very persuasive, reason for the Society to be proud of its willingness to protest human rights violations by the U.S. government.

# Post-1995 APS HR Activities

Many of the human rights activities recounted here, especially those prior to 1995, are based on the annual reports prepared by CIFS, supplemented by reports of small committee accomplishments prepared by CIFS members serving as small committee coordinators. But these just mentioned reports have not been archived, nor would they otherwise be available to this article's typical reader; fortunately, this writer has an adequate supply of such reports. In CIFS's early days, its annual reports were published in the *Bulletin of the APS*; for example, the annual CIFS report for 1984 was published on pp. 1068-72 of the 1985 *Bulletin*. The *Bulletin* is archived on the web only for the years since 1993, however—and this writer was unable to extract any information from the *Bulletin* archive about CIFS Reports, whether dated from before or after 1993.

On the other hand, many interesting articles and Letters to the Editor on the subject of human rights can be found in *Physics Today*, though using the phrases "CIFS Human Rights" and "Human Rights CIFS" to search the *Physics Today* archive produced no hits; the phrases "APS Human Rights" and "Human Rights APS" produced only three hits for articles dated April 1978, July 1985 and June 1995 [29].

For the period after May 1995, however, when both publication and archiving of the *APS News* began, useful information about the Society's human rights activities is much more readily available. The *APS News* archive search of "CIFS" yielded 161 hits, many quite useful. For instance, one of these hits is an article discussing a 2011 CIFS letter to Iran's Grand Ayatollah urging the freeing of Omid Kokabee [30].

In short, the *APS News* archive provides much more useful and accessible material about CIFS human rights activities since 1995 than either *Physics Today* or the *APS Bulletin*. This should not be taken to imply that *APS News* provides as much information about CIFS activities as do the CIFS annual reports; the CIFS annual reports are far more informative. For instance, the CIFS annual report for 2006 describes activities on behalf of individuals in the following countries: Russia, Iraq, Libya, Venezuela, China, the Ukraine and Mexico. The *APS News* for 2006 and 2007 doesn't come close to providing such detailed information.

On the other hand, the last CIFS annual report of which this writer is aware is for the year 2007; apparently, after 2007, the APS Council has not required submission of annual reports by CIFS. Thus for the years 2008 and beyond, the *APS News* is essentially the only useful available source about APS endeavors on behalf of human rights. Use of *APS News* for this purpose is greatly facilitated by the CIFS web page, which has links titled "CIFS Human Rights Cases" and "CIFS in *APS News*". The present (as of July 2015) "CIFS Human Rights Cases" link discusses rights violations in Russia, Gaza, Bahrain and Iran. The Iran discussion states that, in January 2015, Kokabee's retrial upheld his original ten year prison sentence. The same link also informs its readers that in October 2014 Kokabee was also awarded the American Association for the Advancement of Science (AAAS) Scientific Freedom and Responsibility Award. The present "CIFS in *APS News*" link also provides links to a number of *APS News* articles titled "CIFS Briefs" which, as the *APS News* puts it in its July 2014 issue, highlight "the Connection Between Human Rights and Science for the Physics Community." For example, this link carries a story about CIFS support of Sergey Kalakin (hitherto not mentioned in this article), a scientist who is an expert on the safety of nuclear reactors whom the Russian government has accused of embezzlement and fraud [31].

# Other HR Activist Organizations

In the last half century or so the APS has not been the only scientific organization, and physics has not been the only scientific discipline, heavily involved in human rights activities. Such organizations and disciplines include, e.g., the American Association for the Advancement of Science (AAAS), the American Chemical Society (ACS), and the National Academy of Sciences (NAS). Many other organizations, such as the Committee of Concerned Scientists, are composed of scientists who work together on human rights protection though they are not attached to any particular scientific organization. Moreover the AAAS web page titled "Science and Human Rights Coalition" [32], under the heading "Organizations Defending the Human Rights of Scientists," lists 23 such organizations, including the APS and all the other scientific societies mentioned above, but also including, e.g., the New York Academy of Sciences and the American Mathematical Society. The Wikipedia article "List of human rights organizations" [33] includes many human rights organizations, the vast majority non-scientific.

To this writer's knowledge, however, few non-APS human rights organizations have done their essential supportive letter writing via the small committee format CIFS has found to be so effective. Thus this writer believes that, of the scientific organizations involved in human rights activities, the APS has been among the more dedicated and successful.

Although the CIFS is committed to protecting the human rights of all scientists, not merely physicists, a quite significant fraction of the scientists whose human rights the APS has defended have themselves been physicists. This is a remarkable observation, especially considering the comparatively small percentage of physicists in any nation's scientific population. Apparently there is something in the culture of the physics profession—in its insistence on learning how nature truly functions, in its readiness to honor all those who advance this quest no matter what their nationality or the color of their skin—that makes physicists unusually reluctant to quietly accept misuses of state power. All APS members, therefore, should take great pride not only in the American Physical Society's defense of human rights, but also in the inspiring fact that so many of the scientists defended by the APS have been physicists, willing to take actions which can remind future generations of one of the glories of our species, namely that no matter how overwhelming the state power, some humans will refuse to be cowed."

## Acknowledgments

This writer thanks CIFS Administrator Michele Irwin and now retired APS International Affairs Director Irving Lerch for their invaluable assistance in his obtaining CIFS documents, without which assistance this article could not have been written.

**Endnotes**

1. This article is based on a fifty minute March 2015 University of Pittsburgh Physics Department colloquium talk. A thirty minute version of that colloquium talk, condensed because of time constraints, was delivered as an invited talk at the APS Baltimore meeting in April 2015. An even earlier version of the Baltimore talk was delivered at the APS March 2005 meeting in Los Angeles and published in the APS Forum on Physics and Society publication Physics and Society (July 2005). A film of the Baltimore invited talk is available on the web at https://vimeo.com/124204747; it is necessary to enter the password Pitt, using an upper case P.
2. Jewish Telegraph Agency, Dec. 11, 1938, http://www.jta.org/1938/12/11/archive/manifesto.
3. *Am. J. Physics* 68, 595 (2000).
4. *New York Times*, March 5, 1948. This quote was brought to this writer's attention via footnote 116 of Lustig's history, see [3] above.
5. *Physics Today* 29, 103, November 1976.
6. The quote is from the very top of the CIFS page on the APS website, as of July 28, 2015.
7. Undated letter from Parmentola to Kurt Gottfried, (the 1980 CIFS Chair) apparently written on the occasion of a memorial ceremony for 1980 APS President Herman Feshbach, who died in 2000. The assertion that Parmentola served as POPA Executive Director is taken from his bio (obtained via a 2/10/15 googling).
8. The writer is indebted to CIFS Committee Administrator Michele Irwin for providing these illustrations.
9. Bulletin of the American Physical Society, May 1978.
10. John Parmentola, "APS Activities in Human Rights" published in Physics and Society (September, 1979).
11. The term "refusenik" is defined and discussed in the Wikipedia article "Refusenik," accessed 3/11/15.
12. The information about Yuri Orlov's human rights activities in the Soviet Union, and his subsequent punishment by the Soviet authorities, is taken from the Wikipedia article "Yuri Orlov," accessed 3/10/15.
13. The information about Natan Sharansky's imprisonment in the Soviet Union, as well as his later political activities in Israel, is taken from the Wikipedia article "Natan Sharansky," accessed 3/17/15.

14. Bulletin of the APS (1981), p. 683.
15. This CIFS action, which is not mentioned in the 1980 CIFS Annual Report (recall footnote 14 above), was revealed to the writer by Hugh DeWitt himself. DeWitt, who served as CIFS Chair for the year 2000 (and died in 2013 at the age of 83), also told the writer he was convinced the CIFS letter was an important factor in the Lab's eventual decision not to take disciplinary action against him.
16. Posting accessed on 7/28/15 via the CIFS web page. Click on "CIFS Poster".
17. *Physics Today* 34, 27, March 1981.
18. Fang obituary, *Physics Today* 65, 66, September 2012. Like Orlov, Fang received a professorship shortly after arriving in the U.S., at the University of Arizona. Also like Orlov, Fang continued his human rights efforts after emigrating; in particular he chaired CIFS in 1994, and in 1996 was awarded the APS Nicholson Medal for Humanitarian Service.
19. *Physics Today* 34, 51 January 1981.
20. Brailovsky, a computer scientist, was one of the refusenik community's leaders; in particular, the weekly "Sunday Seminar" organized by the refusenik scientists met in Brailovsiky's apartment for an extended period of time starting about 1977. In 1981 he was charged with "anti-Soviet slander" and sentenced to five years of internal exile in Siberia, after a two-day trial in which only his wife and son were admitted to the courtroom. For further details about Brailovsky's refusenik activities and trial, see E. Gerjuoy and D. Arzt, "The Torment of Victor Brailovsky" Human Rights 11, 30, Winter 1983. Human Rights is the Journal of the American Bar Association Section of Individual Rights and Responsibilities.
21. *Physics Today* 38, 71, July 1985.
22. *Physics Today* 42, 83, September 1989.
23. 2014 *APS News*, January, March, May, July and November. The January and July issues discussed Kokabee's receipt in September 2013, jointly with the Russian physicist Boris Altshuler, of the APS Andrei Sakharov human rights Prize. The November issue informed readers that Kokabee had been granted a retrial which, however, did not overturn his sentence. See also footnotes 24-26 below.
24. 2014 *APS News*, January and May. The other organizations included, inter alia, Amnesty International and the Committee of Concerned Scientists.
25. 2014 *APS News*, July.
26. 2014 *APS News*, March.
27. Congressional Record: October 12, 2000, House, pp. H9880-H9894.
28. The Wen Ho Lee case has been the subject of two books: by Stober and Hoffman, *A Convenient Spy* (Simon & Schuster 2001); and by Lee himself with Helen Zia, *My Country Versus Me* (Hyperion 2001).
29. The April 1978 article, by 1977 APS President George Pake, contained instructive information about CIFS early human rights activities; the July 1985 article amplified the CIFS 1984 annual report and included a discussion of CIFS small committee activities; the June 1995 article, a Letter to the Editor written by the 1994 and 1995 CIFS Chairs, concerned Chinese governmental human rights violations.
30. 2011 *APS News* (August/September).
31. 2015 *APS News* (March).
32. As posted 8/1/15.
33. As posted 8/3/15.